\documentclass[sigconf]{acmart}

\AtBeginDocument{%
  \providecommand\BibTeX{{%
    \normalfont B\kern-0.5em{\scshape i\kern-0.25em b}\kern-0.8em\TeX}}}

\copyrightyear{2020} 
\acmYear{2020} 
\setcopyright{acmlicensed}\acmConference[ICSEW'20]{IEEE/ACM 42nd International Conference on Software Engineering Workshops }{May 23--29, 2020}{Seoul, Republic of Korea}
\acmBooktitle{IEEE/ACM 42nd International Conference on Software Engineering Workshops (ICSEW'20), May 23--29, 2020, Seoul, Republic of Korea}
\acmPrice{15.00}
\acmDOI{10.1145/3387940.3392183}
\acmISBN{978-1-4503-7963-2/20/05}

\begin{document}

\title{Towards a Quantum Software  Modeling Language}

\author{Carlos A. P\'erez-Delgado}
\authornote{Both authors contributed equally to this research.}
\orcid{1234-5678-9012}
\affiliation{%
  \institution{University of Kent}
  \city{Canterbury}
  \state{Kent}
  \country{United Kingdom}
  \postcode{CT2 7NF}
}
\email{c.perez@kent.ac.uk}

\author{Hector G. Perez-Gonzalez}
\affiliation{%
  \institution{Universidad Aut\'onoma de San Luis Potos\'i}
  \city{San Luis Potos\'i}
  \state{SLP}
  \country{M\'exico}}
\email{hectorgerardo@uaslp.mx}


\begin{abstract}
  We set down the principles behind a modeling language for quantum software. We present a minimal set of extensions to the well-known Unified Modeling Language (UML) that allows it to effectively model quantum software. These extensions are separate and independent of UML as a whole. As such they can be used to extend any other software modeling language, or as a basis for a completely new language. We argue that these extensions are both necessary and sufficient to model, abstractly, any piece of quantum software. Finally, we provide a small set of examples that showcase the  effectiveness of the extension set.

\end{abstract}



\maketitle

\section{Introduction}

Quantum computation rose to prominence after the discovery of quantum algorithms\cite{shor,grover} that can efficiently perform tasks that are intractable classically. These discoveries propelled research and interest in quantum computation. Today, there exists prototype quantum hardware with computational capabilities beyond that of any classical machine\cite{martinis}. Further applications of quantum theory to computation have also been made in several areas of theory of computing, such as models of computation\cite{luqca}, data structures\cite{quantumgraph}, and cryptography\cite{bb84}.

Quantum computation has, until today, been studied almost exclusively `in the small.' A general understanding of quantum computation, or, quantum programming `in the large' is yet to be developed. Here we aim to set the foundations of a general framework for studying, developing, and conveying quantum programs. We aim to do so by developing a universal modeling language for quantum software. Rather than develop such a language from scratch, we have decided to start from the well-known Unified Modeling Language (UML)\cite{booch}, and introduce a minimum set of extensions that allow it to effectively model quantum software.

 Assuming UML to be a shared common-language upon which we can build, allows us to convey our original extensions much more succinctly.
Our extension set can, however, be applied with little or no modification to any other modeling language. 

\section{Q-UML}

Before discussing in depth the extensions we are introducing, we make a few fundamental observations on which we base the guiding principles for our extension set.

Our first observation is about the nature of quantum computation. The central difference between quantum  and classical computation is in \emph{how} it achieves its goals. Quantum computers have access to quantum algorithms\cite{shor}, and quantum data-structures\cite{quantumgraph}, that are unavailable to classical computers---hence their performance advantage. Algorithms and data-structures are, however, implementation details. Algorithms are an essential design choice while programming in the small. However, they are more often than not completely ignored in large-scale software architectural design. For instance, UML diagrams seldom portray algorithms and data-structures beyond a very high-level design perspective.

It would seem then that quantum computation introduces nothing to computation that needs to be captured in a software design diagram. This is not the case, and the reason for this is our second observation. Quantum computation changes the very nature of \emph{information} itself. Quantum information is much richer than classical information. It is also much more challenging to store, transmit, and receive. If a module (class, object, \emph{etc.}) needs to store, transmit or receive quantum information, then this is an important design consideration---which needs to be included in any effective software design.

A third observation here is that the classical vs. quantum nature of the information used by a module is an important consideration both when discussing its internal implementation \emph{and} its interface. Furthermore, these two are separate and independent considerations. 

A classical module, implementing some classical behavior, would have no need, or capability, to communicate quantum data. A quantum module may or may not have to; \emph{i.e.} a module's quantum behavior may be completely part of its internal implementation and not appear as part of its interface. For instance, take a module implementing Shor's algorithm. Shor's algorithm uses quantum effects to efficiently factor a large integer into its prime factors. The implementation of this module must necessarily be quantum. Both the input (the large integer) and the output (the prime factors), consist of classical information. And hence, the \emph{interface} of such a module can be strictly classical. 

More generally, we can conceive of quantum software modules that have all classical inputs and outputs (like the above example), all quantum inputs and outputs, or a mix of both. A quantum software design must address, for each individual interface element, whether it is classical input/output, or if it is quantum. 
In short, whether a module communicates classically or via quantum information, and whether its internal implementation requires quantum hardware are important considerations that need to be captured in a design document.

The importance of such labelling should be clear. Quantum data can only be stored and transmitted with special hardware designed to do so. More importantly, from an abstract, device-independent, strictly software perspective: quantum and classical information are \emph{not} interchangeable. Classical information is clone-able and admits fanout operations, while quantum information (in general) does not. On the other hand, quantum information has a much larger state-space.

Finally, it is true that quantum information is strictly a super-set of classical information---and hence a quantum module can communicate any classical information it desires using a quantum interface element. We argue, however, that using a quantum interface element and messaging when classical would suffice is bad quantum software design, for the reasons stated above.

In summary, the guiding principles behind any quantum software modeling language must include the following:

\begin{enumerate}
\item \textbf{(Quantum Classes):} Whenever a software module makes use of quantum information, either as part of its internal state/implementation, or as part of its interface, this must be clearly established in a design document.
\item \textbf{(Quantum Elements):} Each module interface element (\emph{e.g.} public functions/methods, public variables) and  internal state variables can be either classical or quantum, and must be labelled accordingly.
\begin{enumerate}
\item \textbf{(Quantum Variables):} Each variable should be labelled as classical or quantum. If the model represents data types, the variables should also specify the classical (\emph{e.g.} integer, string) or quantum (\emph{e.g.} qubit, qubit array, quantum graph state) data type,
\item \textbf{(Quantum Operations):} For each operation, both the input and output should be clearly labelled as either classical or quantum. Whether the operation internally operates quantumly should also be labelled.
\end{enumerate}
\item \textbf{(Quantum Supremacy):} A module that has \emph{at least} one quantum element is to be considered a quantum software module, otherwise it is a classical module. Quantum and classical modules  should be clearly labelled as such.
\item \textbf{(Quantum Aggregation):} Any  module that is composed of one or more quantum modules will itself be considered a quantum module, and must be labelled as such.
\item \textbf{(Quantum Communication):} Quantum and classical modules can communicate with each other as long as their interfaces are compatible, \emph{i.e.} the quantum module has classical inputs and/or outputs that can interface with the classical module.
\end{enumerate}

We will argue in Sec.\ \ref{sec:discussion} how these extensions are not only necessary, but also sufficient in order to design and represent quantum software. First, in the following two sections we put these principles into practice as a set of concrete extensions to UML. 

\subsection{Class Diagram Extensions}

UML is a very graphical language, meant to convey a lot of meaning in a very small amount of space. As such, it makes sense to use a graphical way to represent quantum software elements. We chose to do this by use of \textbf{bold} text to denote quantum elements, and double lines to denote a quantum relationship or quantum communication.

\begin{figure}[th]
  \centering
  \includegraphics[width=\linewidth]{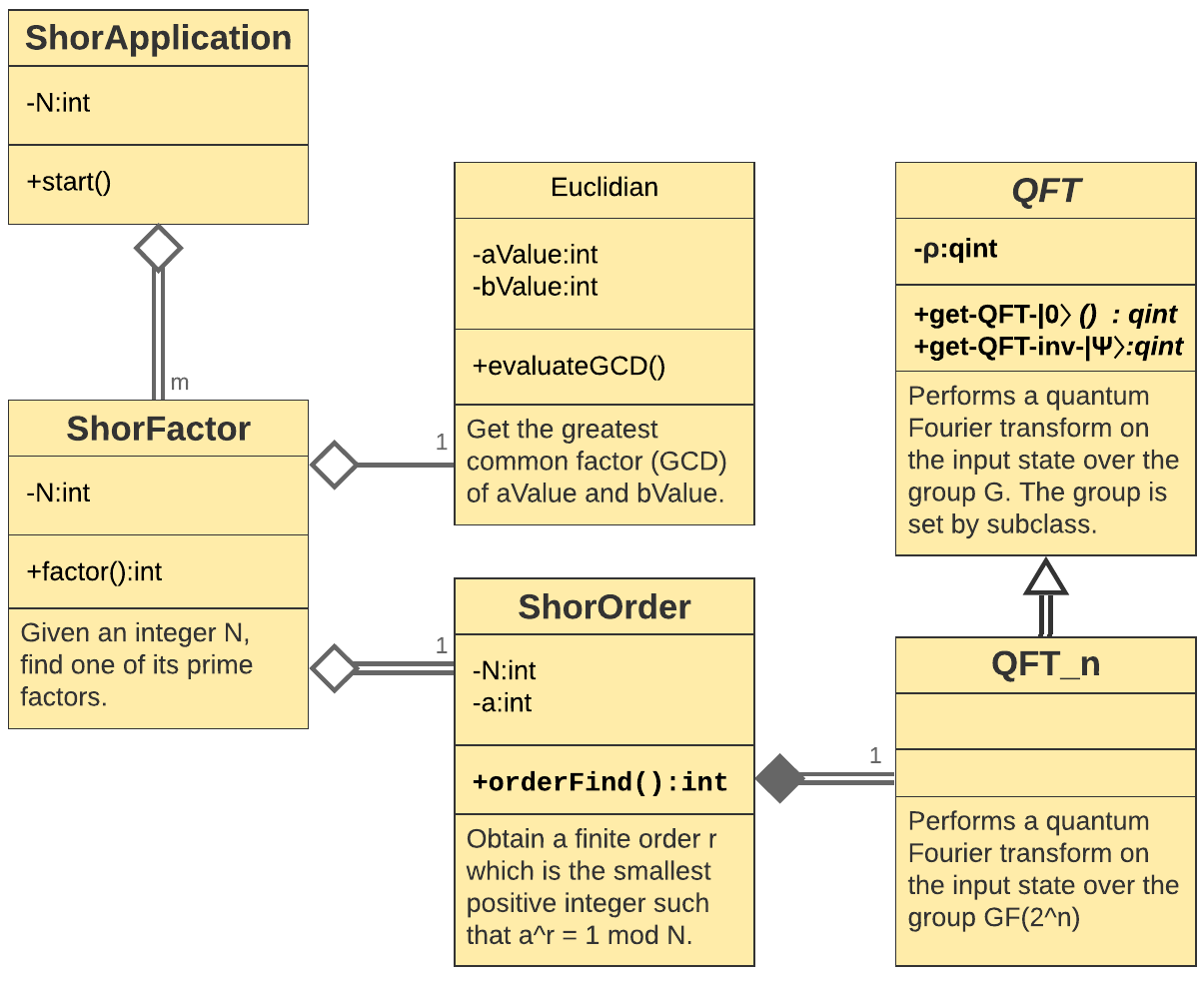}
  \caption{Q-UML class diagram of Shor's Algorithm. Quantum classes and interface elements are presented in bold text, and quantum relationships use double-lines.}\label{fig:class}
  \Description{Q-UML Class Diagram}
\end{figure}

For attributes, the name will be bold if it is represented using quantum information. For methods, we use the following convention. If any of the inputs are quantum, these are bold. If the output or datatype of the method is quantum, then the datatype should also be bold. For backwards compatibility with regular UML, whenever the input or output datatypes of a method are omitted, these will be assumed to be classical in nature. If a class/object has any quantum attributes or methods then it itself is considered quantum, and its name shall also be bold.

Relationships between classes will use double-lines whenever the relationship is quantum in nature. For inheritance, if the superclass is quantum then the subclass, and the inheritance relationship, will also be quantum.  (the converse is not necessarily true however). In the case of aggregation and composition, if a class/object being aggregated/composed is quantum, then the class/object to which it is aggregated/composed into, as well as that relationship will also be quantum. Association relationships do not have any special rules, beyond the need of a quantum class/object to have a classical interface if it is to associate with classical classes/objects.

Fig.\ \ref{fig:class} showcases a Q-UML diagram that exemplifies the above rules.

\subsection{Sequence Diagram Extensions}

Sequence diagrams in UML allow us to portray the \emph{dynamic} relationship between modules in a software program. As we did before for static relationships, we extend the existing language in order to allow us to differentiate between classical and quantum messages. As previously discussed, this is essential information. Quantum information behaves differently from classical information; it can store/portray different data; it admits different operations; and, it requires different hardware to store, send, and receive. 

\begin{figure}[th]
  \centering
  \includegraphics[width=\linewidth]{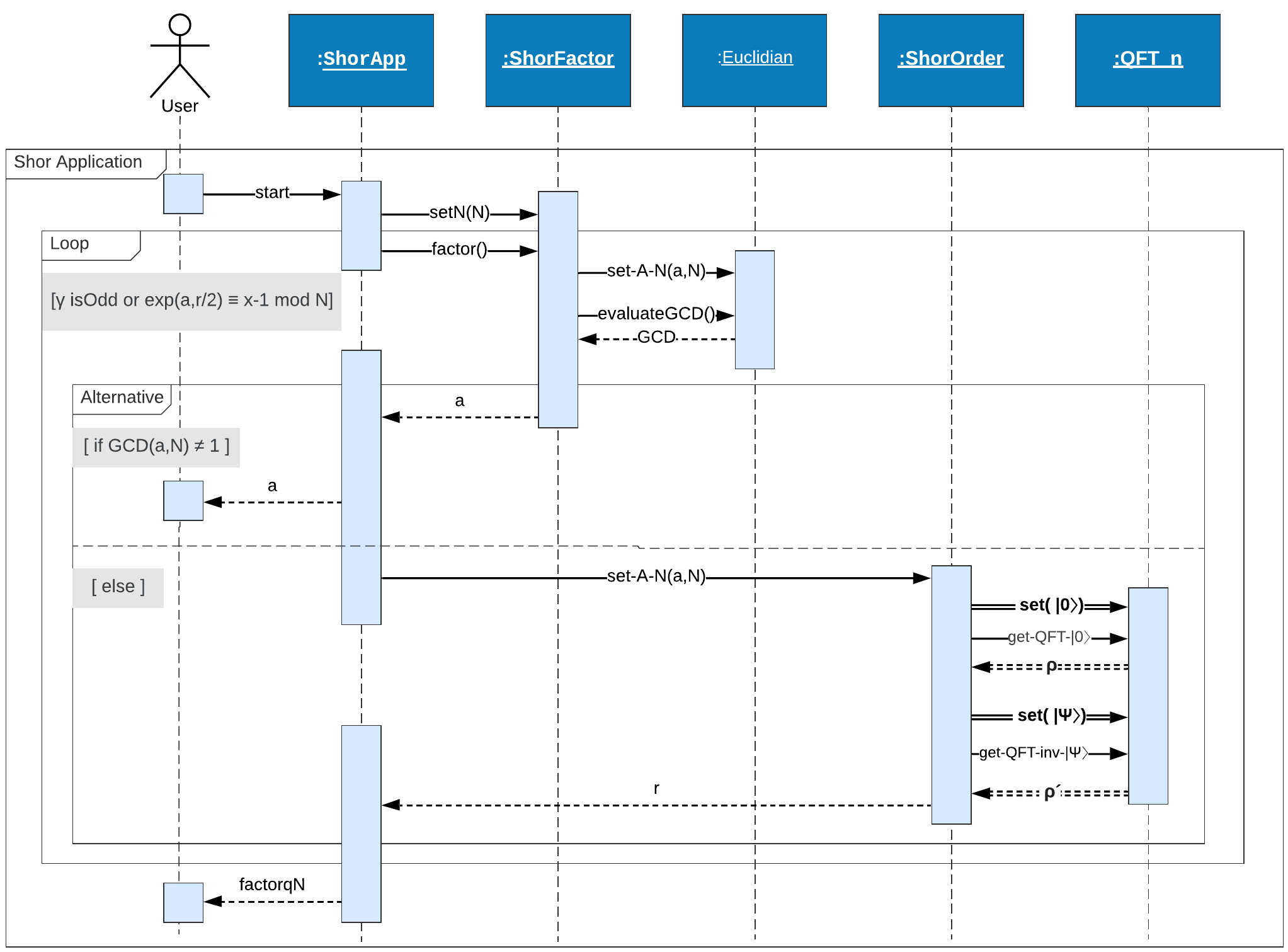}
  \caption{Q-UML sequence diagram of Shor's Algorithm. Quantum classes are presented in bold text, and quantum messages use double-lines.}\label{fig:sequence}
  \Description{Q-UML Sequence Diagram. }
\end{figure}

Like before, we make use of \textbf{bold} text to markup quantum modules, and double lines to portray quantum messages. Fig.\ \ref{fig:sequence} shows a Q-UML sequence diagram. Note how even though the \emph{relationship} between \textbf{Shorfactor} and  \textbf{ShorOrder} is quantum, the messaging between them is \emph{not.} This illustrates an important point. A module is marked as quantum if it uses quantum resources in any form, either directly as part of its internal implementation or as part of an aggregated module. If a sub-module (in UML a \emph{composed} class or object) is quantum, then the encompassing module must also be marked as quantum. In a static (\emph{e.g.} class) diagram, the quantum composition relationships inform us---especially in the case of a seemingly classical module that does not \emph{in itself} use quantum resources---which composed modules are using quantum resources.

Also, note the communication between the objects  \textbf{ShorOrder} and  \textbf{QFT\_n}. The module  \textbf{QFT\_n} operates on a quantum state. Hence, both `set' messages are quantum. Likewise, the return messages $\rho$ and $\rho'$ are quantum states. However, the request to perform a quantum Fourier transform (QFT) or a QFT inverse operation can (and therefore should) be communicated classically. This diagram showcases the level of granularity available to us using these diagrams with the proposed extensions.

\subsection{Discussion}\label{sec:discussion}

We have proposed a minimal series of extensions to existing software modeling languages. We exemplify our additions in UML, but these extensions are easily applicable to any other modeling language, or be used as the basis for a new modeling language. 

We've argued the necessity of each of the extensions in previous sections. We can argue as well, that these extensions are not only necessary, but also sufficient to fully model quantum software. To make this argument, we appeal to the fact that all quantum computation is simulable using classical computation albeit with an efficiency loss. Other than their use of quantum information and algorithms, quantum computers are indistinct from classical ones. Hence, from a high-level design perspective, the only information element that needs to be considered when developing quantum software is when quantum (rather than classical) information is being used. 

The one remaining information element we have not discussed is algorithm efficiency. If quantum computation is to be used, it will most likely be due to the efficient algorithms at its disposal. That said, algorithm efficiency is not a solely quantum consideration. UML itself does not inherently have language elements for algorithm efficiency (beyond user-defined notes). It does, however, have several extensions used and proposed for this purpose(see \emph{e.g.}\cite{UMLperformance}). Other modeling languages may also have definite algorithm efficiency elements. We argue that it is best to use existing language elements when they are available.

\begin{acks}
CP-D would like to acknowledge funding through the EPSRC Quantum Communications Hub (EP/T001011/1).
The authors would also like to thank Joanna I. Ziembicka for useful comments during the preparation on this manuscript.
\end{acks}

\bibliographystyle{ACM-Reference-Format}
\bibliography{bibliography}


\begin{thebibliography}{8}


\ifx \showCODEN    \undefined \def \showCODEN     #1{\unskip}     \fi
\ifx \showDOI      \undefined \def \showDOI       #1{#1}\fi
\ifx \showISBNx    \undefined \def \showISBNx     #1{\unskip}     \fi
\ifx \showISBNxiii \undefined \def \showISBNxiii  #1{\unskip}     \fi
\ifx \showISSN     \undefined \def \showISSN      #1{\unskip}     \fi
\ifx \showLCCN     \undefined \def \showLCCN      #1{\unskip}     \fi
\ifx \shownote     \undefined \def \shownote      #1{#1}          \fi
\ifx \showarticletitle \undefined \def \showarticletitle #1{#1}   \fi
\ifx \showURL      \undefined \def \showURL       {\relax}        \fi
\providecommand\bibfield[2]{#2}
\providecommand\bibinfo[2]{#2}
\providecommand\natexlab[1]{#1}
\providecommand\showeprint[2][]{arXiv:#2}

\bibitem[\protect\citeauthoryear{Arute~\emph{et. al.}}{Arute~\emph{et.
  al.}}{2019}]%
        {martinis}
\bibfield{author}{\bibinfo{person}{Frank Arute~\emph{et. al.}}}
  \bibinfo{year}{2019}\natexlab{}.
\newblock \showarticletitle{Quantum supremacy using a programmable
  superconducting processor}.
\newblock \bibinfo{journal}{\emph{Nature}} \bibinfo{volume}{574},
  \bibinfo{number}{7779} (\bibinfo{year}{2019}), \bibinfo{pages}{505--510}.
\newblock
\showISBNx{1476-4687}
\urldef\tempurl%
\url{https://doi.org/10.1038/s41586-019-1666-5}
\showDOI{\tempurl}


\bibitem[\protect\citeauthoryear{Bennett and Brassard}{Bennett and
  Brassard}{2014}]%
        {bb84}
\bibfield{author}{\bibinfo{person}{Charles~H Bennett} {and}
  \bibinfo{person}{Gilles Brassard}.} \bibinfo{year}{2014}\natexlab{}.
\newblock \showarticletitle{Quantum cryptography: public key distribution and
  coin tossing.}
\newblock \bibinfo{journal}{\emph{Theor. Comput. Sci.}} \bibinfo{volume}{560},
  \bibinfo{number}{12} (\bibinfo{year}{2014}), \bibinfo{pages}{7--11}.
\newblock


\bibitem[\protect\citeauthoryear{Booch, Rumbaugh, and Jacobson}{Booch
  et~al\mbox{.}}{2005}]%
        {booch}
\bibfield{author}{\bibinfo{person}{Grady Booch}, \bibinfo{person}{James
  Rumbaugh}, {and} \bibinfo{person}{Ivar Jacobson}.}
  \bibinfo{year}{2005}\natexlab{}.
\newblock \bibinfo{booktitle}{\emph{Unified Modeling Language User Guide, The
  (2nd Edition) (Addison-Wesley Object Technology Series)}}.
\newblock \bibinfo{publisher}{Addison-Wesley Professional}.
\newblock
\showISBNx{0321267974}


\bibitem[\protect\citeauthoryear{{Canevet}, {Gilmore}, {Hillston}, {Prowse},
  and {Stevens}}{{Canevet} et~al\mbox{.}}{2003}]%
        {UMLperformance}
\bibfield{author}{\bibinfo{person}{C. {Canevet}}, \bibinfo{person}{S.
  {Gilmore}}, \bibinfo{person}{J. {Hillston}}, \bibinfo{person}{M. {Prowse}},
  {and} \bibinfo{person}{P. {Stevens}}.} \bibinfo{year}{2003}\natexlab{}.
\newblock \showarticletitle{Performance modelling with the Unified Modelling
  Language and stochastic process algebras}.
\newblock \bibinfo{journal}{\emph{IEE Proceedings - Computers and Digital
  Techniques}} \bibinfo{volume}{150}, \bibinfo{number}{2}
  (\bibinfo{date}{March} \bibinfo{year}{2003}), \bibinfo{pages}{107--120}.
\newblock
\showISSN{1350-2387}
\urldef\tempurl%
\url{https://doi.org/10.1049/ip-cdt:20030084}
\showDOI{\tempurl}


\bibitem[\protect\citeauthoryear{Grover}{Grover}{1996}]%
        {grover}
\bibfield{author}{\bibinfo{person}{Lov~K. Grover}.}
  \bibinfo{year}{1996}\natexlab{}.
\newblock \showarticletitle{A Fast Quantum Mechanical Algorithm for Database
  Search}. In \bibinfo{booktitle}{\emph{Proceedings of the Twenty-eighth Annual
  ACM Symposium on Theory of Computing}} \emph{(\bibinfo{series}{STOC '96})}.
  \bibinfo{publisher}{ACM}, \bibinfo{address}{New York, NY, USA},
  \bibinfo{pages}{212--219}.
\newblock
\showISBNx{0-89791-785-5}
\urldef\tempurl%
\url{https://doi.org/10.1145/237814.237866}
\showDOI{\tempurl}


\bibitem[\protect\citeauthoryear{P\'erez-Delgado and Cheung}{P\'erez-Delgado
  and Cheung}{2007}]%
        {luqca}
\bibfield{author}{\bibinfo{person}{Carlos~A. P\'erez-Delgado} {and}
  \bibinfo{person}{Donny Cheung}.} \bibinfo{year}{2007}\natexlab{}.
\newblock \showarticletitle{Local unitary quantum cellular automata}.
\newblock \bibinfo{journal}{\emph{Phys. Rev. A}}  \bibinfo{volume}{76}
  (\bibinfo{date}{Sep} \bibinfo{year}{2007}), \bibinfo{pages}{032320}.
\newblock
Issue 3.
\urldef\tempurl%
\url{https://doi.org/10.1103/PhysRevA.76.032320}
\showDOI{\tempurl}


\bibitem[\protect\citeauthoryear{Shor}{Shor}{1994}]%
        {shor}
\bibfield{author}{\bibinfo{person}{Peter~W Shor}.}
  \bibinfo{year}{1994}\natexlab{}.
\newblock \showarticletitle{Algorithms for quantum computation: Discrete
  logarithms and factoring}. In \bibinfo{booktitle}{\emph{Proceedings 35th
  annual symposium on foundations of computer science}}. Ieee,
  \bibinfo{pages}{124--134}.
\newblock


\bibitem[\protect\citeauthoryear{Zhao, P\'erez-Delgado, and Fitzsimons}{Zhao
  et~al\mbox{.}}{2016}]%
        {quantumgraph}
\bibfield{author}{\bibinfo{person}{Liming Zhao}, \bibinfo{person}{Carlos~A.
  P\'erez-Delgado}, {and} \bibinfo{person}{Joseph~F. Fitzsimons}.}
  \bibinfo{year}{2016}\natexlab{}.
\newblock \showarticletitle{Fast graph operations in quantum computation}.
\newblock \bibinfo{journal}{\emph{Phys. Rev. A}}  \bibinfo{volume}{93}
  (\bibinfo{date}{Mar} \bibinfo{year}{2016}), \bibinfo{pages}{032314}.
\newblock
Issue 3.
\urldef\tempurl%
\url{https://doi.org/10.1103/PhysRevA.93.032314}
\showDOI{\tempurl}


\end{thebibliography}
\end{document}